# Scholarly outputs of EU Research Funding Programs:

*Understanding differences between datasets of publications reported by grant holders and OpenAIRE Research Graph in H2020*

Alexis-Michel Mugabushaka[1], Miriam Baglioni[2], Alessia Bardi[2], Paolo Manghi[3]

[1] European Commission, DG RTD, Unit G2[1]
[2] Istituto di Scienza e Tecnologie dell'Informazione, CNR, Italy
[3] OpenAIRE A.M.K.E

**Abstract**

Linking research results to grants is an essential prerequisite for an effective monitoring and evaluation of funding programs. For the EU research funding programs, there are multiple datasets linking scholarly publications to the individual grants, including both open data and those from commercial bibliometric databases. In this paper, we systematically compare openly available data from two data sources: on one hand those reported by the Grant holders (and subsequently published by the European Commission on open data portal) and those from the OpenAIRE Research Graph which collect data from multiple sources. We describe the dataflow leading to their creation and assess the quality of data by validating, on sample basis, the link <project, publications>. We report that, by and large, OpenAIRE Research Graph offers a more complete dataset of scholarly outputs of from EU Research funding programs. We identify also possible improvements and make recommendations on how they can be addressed.

## 1. Introduction

The potential of bibliometrics to inform evaluation of European Union Research Funding Programs (FP) have been known for decades. A meta-analysis of evaluative activities from mid-1980ies to 1995 noted that "*the scientific and technical quality of the research*" has been a constant evaluation criterion that called for a "*continuous monitoring, by independent external experts aided by output indicators*" [Georghiou, L. (1995)]. The effective use of bibliometrics has been however hindered by lack of adequate data especially publications linked to specific projects. Illustratively: a review paper of evaluation activities of the FPs lists a variety of methods used (in period 1999 to mid-2004) in which bibliometrics is conspicuous by its absence [Arnold et al. (2005)]. In evaluation of FP6, there have been some attempts to apply bibliometric analysis, but they remain confined to specific aspects and do not attempt to apply systematically bibliometrics analysis to a wide range of questions [Vanecek(2010)] and the official report of the evaluation of the 6th Framework Programme makes no references to publications coming from projects [EC 2009]. In 2009, Health Research for Europe (HR4E), a project which was funded with the aim of analyzing health-related research funded by the Framework Programs (FP5 and FP6), noted the lack of results data and publication data in particular. In its recommendations, it calls for "*EU to place at least as great an emphasis on the outputs of the research as is presently placed on the inputs*." [2]

With respect to the availability of publication data, the situation has changed in meantime. Currently, researchers and other stakeholders interested in the analysis of scholarly results of EU funded projects have the choice between different sources including data provided by European Commission (EC) notably on the open data portal[3], datasets from the OpenAIRE

---

[1] The views expressed in this paper are the authors. They do not reflect the views or official positions of the European Commission

[2] https://cordis.europa.eu/project/id/37397/reporting

[3] https://data.europa.eu/data/

Research Graph and from some of the commercial bibliographic databases. From the perspective of the users, the challenge is less to access the data than to choose among those sources or assess the differences. A better understanding of the commonalities and differences between the different data sources can increase the transparency of evaluative studies of EU funding programs and help make sense of discrepancies, inevitable if different data sources are used.

The objective of this paper is to systematically compare publication data from two openly available sources: on one hand those reported by grant holders to the EC and on the other hand, the OpenAIRE datasets. Focusing on H2020 Program, the paper documents the dataflows which lead to the creation of the two open datasets and gauge their differences.

The next section describes various data sources for publications from EU projects. It is followed by a detailed analysis of the differences between reported publications and those found in the OpenAIRE Research Graph. The last, fourth section summarizes our findings and offers some reflections on how the curation of the scholarly outputs from EU funding can be improved.

## 2. Data sources for scholarly results of EU funded projects

For analytical purposes, we distinguish here between three main data sources: (1) publications reported by grant holders to the EC, (2) OpenAIRE Research Graph and (3) commercial bibliometric databases. However, as we explain in the following section, the boundaries between those categories are blurred as some of those sources are also intertwined in terms of their primary data sources and data collection mechanisms. We first briefly describe the reported publications and OpenAIRE Research Graph dataset, then the commercial bibliometric databases.

*2.1. Reported publications and OpenAIRE dataset*

Beginning from the $7^{th}$ Framework Programmes, the EC started collecting systematically publications to which the funding has fully or partially contributed. At reporting time (usually twice in course of the project), the reporting of the publications was done by grant holders who entered metadata of publications in a dedicated structured form. Details on the data management and curation efforts for the publications collected during the $7^{th}$ FP are documented in a previous work on "*linking Publications to Funding at Project Level*" which focused on the $7^{Th}$ Framework Program (Mugabushaka, 2020).

For H2020, a new system for Grant Management (SyGMa) was developed which changed this process. According to the H2020 Online Manual[4], publications can be reported at any time during the project and several features are foreseen to make it easier for users and enhance the quality of collected data:

- The reporting form is pre-filled with publications linked to the projects retrieved via OpenAIRE. The Grant holder is asked to "*Simply check if the references are directly linked to the work performed within the project. If so, tick these publications as relevant and they will then be included in the table of publications when the report is generated*."
- For publications not suggested via OpenAIRE, grant holders are asked to enter a Digital Object Identifier (DOI) and the metadata are retrieved from Crossref.
- The Grant holders are asked to enter manually metadata of publications which are neither in OpenAIRE nor in Crossref (i.e., not retrievable through the DOI).

OpenAIRE (www.openaire.eu) is a pan-European infrastructure for research. It started in 2006 as an initiative to network Open Access repositories (funded as project DRIVER) and has evolved into a full-fledged infrastructure to support the European Union Open Science Agenda. It has created, among others, an open and participative infrastructure linking scientific publications to funding, research data and other entities of the research life cycle (e.g., authors with ORCiD, organisations, research software).[5]

---
[4] https://ec.europa.eu/research/participants/docs/h2020-funding-guide/index_en.htm

In terms of linking publications to grants, OpenAIRE uses a standardized data model, which also includes a source of funding. By harvesting data from repositories which follow its guidelines[6], OpenAIRE is able link the publications to specific funding grants. In addition, OpenAIRE also retrieves funding data in structured form from open bibliographic systems notably CrossRef and EuropePMC. Full-texts of Open Access publications are processed by full-text mining algorithms in order to identify acknowledgements to project grants. The precision of the mining algorithm for EC FP7 and H2020 projects is 99.5% . Recall is higher than 99% when projects are properly acknowledged using project/grant IDs. These rates are based on tests performed by the OpenAIRE mining team with manual validation of the results. The last method to gather the links between publications and funding grants is the Link functionality available the OpenAIRE portals[7]. This functionality allows authenticated users to "claim" the work and link it to the funding.

Every three months, SyGMa reports back to OpenAIRE the list of publications that have been reported (those suggested by OpenAIRE and confirmed, those retrieved by SyGMa via Crossref, and those manually entered by grant holders).

This new workflow effectively links the SyGMa reporting system to OpenAIRE as shown in figure 1.

*Figure 1: The reporting workflow in H2020.*

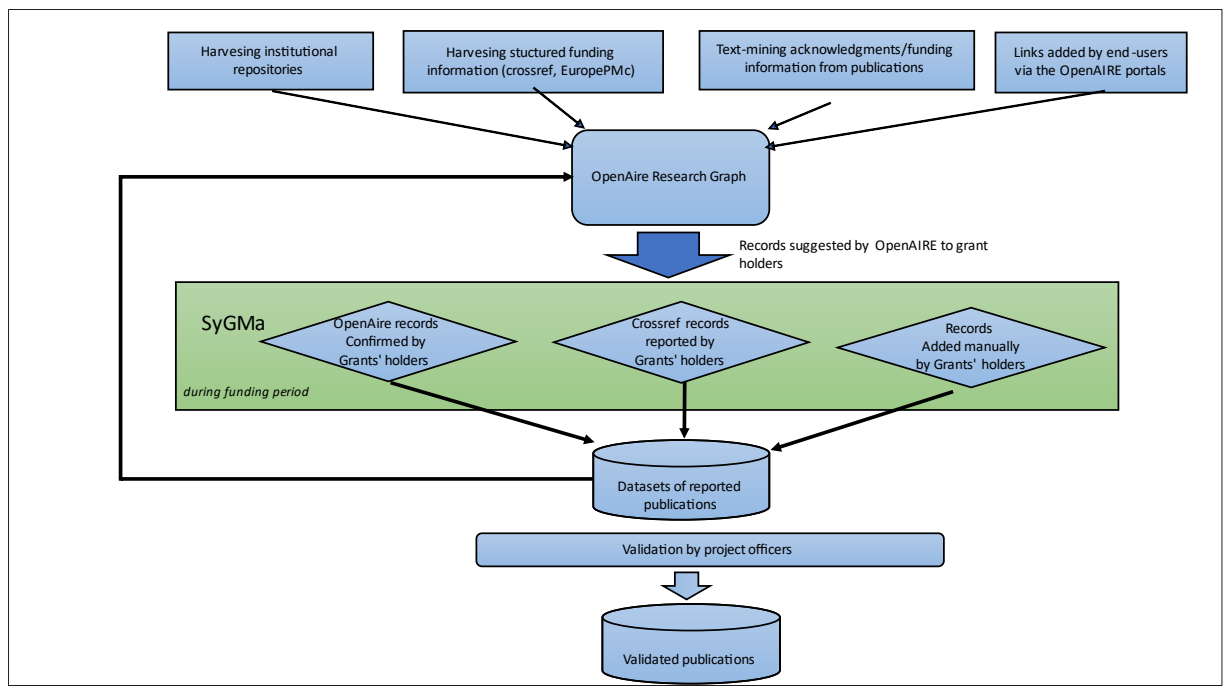

---

[5] https://graph.OpenAIRE.eu/

[6] OpenAIRE guidelines: www.guidelines.OpenAIRE.eu

[7] OpenAIRE. How to report your publication and data to the EC. https://www.OpenAIRE.eu/reporting-to-the-ec

*2.2. EU funding in commercial bibliometric databases*

To our knowledge, there are currently three commercial bibliometric databases in which data on EU funding can be found (1) Web of Science (2) Scopus and (3) Dimensions, all of which are based on proprietary datasets, whose usage is regulated by their respective licenses.

To get more information on how EU funding is linked to publications, we reached out to the corresponding authors of papers Birkle et al. (2020), Baas et al. (2020) and Herzog et al (2020) describing those databases in the special issues of the Quantitative Science Studies (issue 1, vol 1, 2020) and received – in personal communications[8] – the following information:

- The *Web of Science* (WoS) operated by Clarivate started including funding acknowledgements in 2008. The system captures English-language statement containing funding information ("funding statement") which is located typically in the acknowledgements section. The database provides this "funding statement text" alongside the structured grant agency names and grant numbers which are extracted from it. Within the Web of Science, different collections have different coverage, but the acknowledgement extraction covers mostly all document types indexed. Given that all not published items do have funding acknowledgements, and authors may not include all grants, the Web of Science started in late 2020 also to collect additional grant information from several sources. In personal communication the following sources are listed: the US Federal Reporter, the Japanese Database of Grants-in-Aid for Scientific Research (KAKEN), Medline, NIH, NSF, and the Researchfish system used by several UK funding bodies.
- Elsevier, which operates Scopus, has published an extensive documentation of the process by which it identifies potential references to funding acknowledgements [Elsevier 2020]. It reports on an analysis conducted on about 3 Mio. awarded grants by a number of large funding agencies which found that only 45% had publications linked to them. The finding "suggests that the funding information collected by funding agencies is far from complete and needs to be combined with other sources." They also observed that "the information provided by authors via this structured way is not only incomplete, but it also includes a significant number of omissions and errors." The white paper describes a full automated, machine learning-based, process for the extraction of structured funding information from articles. The data pipeline makes also uses of Funders' Registry, an open-source funder's taxonomy developed and maintained by Elsevier and published currently by Crossref. The solution has been patented and its algorithm published [Kayal et al. 2017]. In addition to extracting funding information directly from full text body of the articles, Scopus complements that with data reported to funding agencies by authors.
- Dimensions from Digital Science, is a relatively new bibliometric database, having started only in 2018. It sources data from several organizations. Systems such as Crossref and PubMed Central serve as the "backbone" for publication data. Those data are enriched with data derived from full-text articles and books to add metadata, citations, funding acknowledgements etc. According to information we received in personal communication, funding data in Dimensions are collected on one hand by mining acknowledgments and funding statements in full text articles and on other hand funding data come directly from the funding organisations, especially when they are Dimensions partners. According to Dimensions, its distinctive feature is that it includes not only grant identifiers but also other funding information. Dimensions cover over 6000 funders of which 600+ funders are also indexed to include grants details such as Principal

---

[8] the information was provided by Mrs Tilla Edmunds (Clarivate), Mr Cristina Huidiu (Digital Science) and Mr Attila Emecz (Elsevier)

investigators and co-investigators, funding amounts, grant number … The list of sources was not disclosed to us as it is available only to subscribers.

In a second iteration we also asked them whether their respective systems indexes publication funding information from open datasets which provide links between publications and EU Union grants. The replies we received as shown in the table 1 below.

*Table 1: Indexing of open datasets of publication funding information by commercial bibliometric databases*

| Bibliometric Database | includes publication funding information from the European Open Data Portal | | | includes publication funding information from OpenAire | | | includes publication funding information from Crossref | | |
|---|---|---|---|---|---|---|---|---|---|
| | Yes | No | Prefer not to tell | Yes | No | Prefer not to tell | Yes | No | Prefer not to tell |
| Scopus | | x | | x | | | | x | |
| Web of Science | | x | | | x | | | x | |
| Dimensions | | x | | | x | | x | | |

The table shows major differences between the commercial bibliometric databases (in terms of open data sources). This (together with differences in coverage, Visser 2021) could explain discrepancies in results of studies conducted on different databases. A systematic comparison of the platforms (to each other and with the open datasets) falls outside the scope of this paper and is left to future work.

We note also that publications linked to EU funding can also be found in non-commercial bibliographic systems, which are not considered here. An example is the EuropePMC database which, for the EU funding programmes, includes grant information only for European Research Council (ERC) Life Science projects. Another is Crossref, which provide also grant information provided by publishers. Both sources are used by OpenAIRE.

## 3. Differences between Reported publications and OpenAIRE Research Graph

In the following we describe our work to compare the datasets of publications reported by grant holders and those included in OpenAIRE Research Graph.

In this context a record refers to a pair <project, publication> because a project funding can refer to several publications and a publication can have funding from several projects contributing to it.

We create a record identifier by concatenating the project identifier and a doi, whereby, for practical purposes, we restrict the set only to doi from Crossref. An important implication of this restriction is that the comparison is only partial as it does not cover publications in outlets not indexed in Crossref. But this approach guarantees a comparability of the two datasets by relating both of them to an external reference. We consider that a record is matched in both datasets if the concatenated identifier is present in both dataset and not matched otherwise.

The comparison is based on the following datasets:

1. SYGMA: reported publications as of 2 December 2020
2. OpenAIRE: H2020 publications available in the OpenAIRE Research Graph as of November 2020. It includes:
    - ☐ List of H2020 projects updated on 24 November 2020

- Publications reported in SyGMa until September 2020 (communicated to OpenAIRE via the workflow described in the previous section)
- Links project-publication from Crossref updated on 2 October 2020

Figure 2 shows the share of records compared – for each dataset and the results of the matching process.

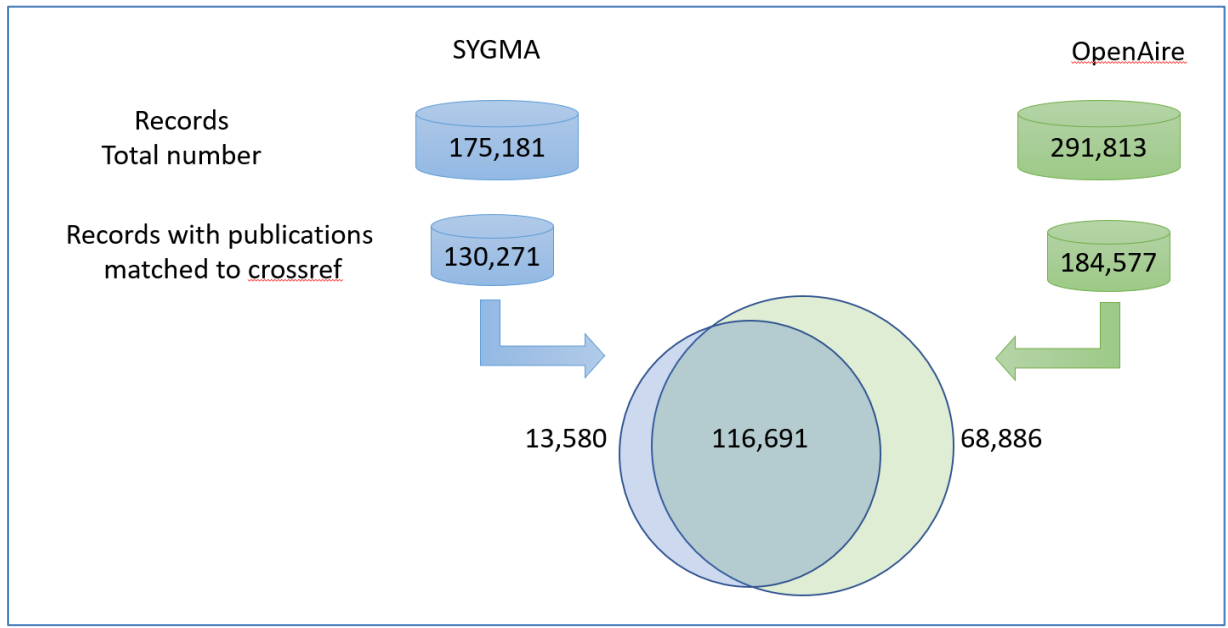

**Figure 2: Overview of comparisons of the two datasets**

In terms of pairs compared in both datasets we have 130k in SyGMa and 185k in OpenAIRE. The comparison shows about 14k records that are in SyGMa but not in OpenAIRE and about 5 times that many in OpenAIRE but not in SyGMa.

In the next sections we take a look at each of those differences separately to understand the mismatch, before discussing the implications of our findings.

*3.2. Records in SyGMa but not in OpenAIRE*

Figure 2 shows there are 13,580 records (links publication – project) available in SyGMa but not in OpenAIRE. Considering the workflow in place for data exchange between SyGMa and OpenAIRE, these missing records are unexpected, and we applied the workflow in Figure 3 to investigate.

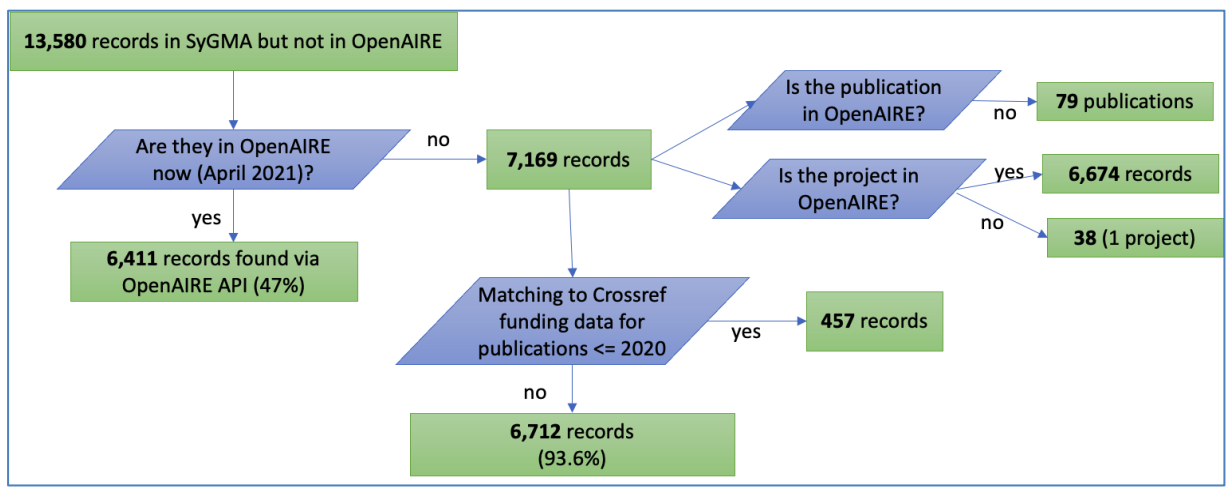

**Figure 3: Verification of records in SyGMa but not in OpenAIRE**

Considering that the analyzed OpenAIRE dataset included input from SyGMa updated at September 2020, we started by checking if the missing links entered in OpenAIRE with a delay. By querying the OpenAIRE API in April 2021, we were able to find about the 47% of the missing links (6,411). For the remaining missing links (7,169), we checked if the linked objects are available in OpenAIRE. All the linked projects are available but one, which started in December 2020 and has in the SyGMa dataset 38 reported publications. As for the publications, they are also all available except 79. The missing 79 publications can be explained by the fact that the OpenAIRE Research Graph version of April includes Crossref updated in December (so, the most recent publications were not yet included). This indicates that most publications are available via OpenAIRE but without the link to the project. The investigation therefore proceeded to understand why the links were not collected by OpenAIRE via Crossref or by processing the reports that SyGMa sends every three months.

The publications involved in the 7,169 missing links are in OpenAIRE and were collected from Crossref: if Crossref has the links to the H2020 projects, those should have been mapped as well. Until April 2021, the rules to map links to H2020 projects from Crossref are:

- The funder DOI is one of:
    - 10.13039/100010663 (H2020 ERC)
    - 10.13039/100010661 (H2020 Programme)
    - 10.13039/501100007601 (H2020)
    - 10.13039/100010665 (H2020 Marie Curie Actions)
    - 10.13039/501100000780 (EC, may include also links to FP7 projects)
    - 10.13039/501100000781 (ERC, may include also links to FP7 projects)
- The funder name matches:
    - "European Union's Horizon 2020 research and innovation program"
    - "European Union's" (may include also links to FP7 projects)

Since we could not map the links with the above rules, we tried to relax the rules and use a more inclusive approach by checking if the funder name contains one of: "ERC", "ERA", "ICT", "CSIC", "Curie", "FET", "European", "EU", "EC", "H2020", "Horizon 2020, "Horizon2020". We adopted the relaxed rules on publications published before 2021 and found out that over the 7,169 missing links, only 457 could have been retrieved. In other words, 93,6% of links not available in OpenAIRE are also not available in Crossref.

A sample-based inspection of some of those records show that over a half belong to documents type which typically do not have acknowledgement of funding (at least not in easily extractable format) such as book chapters and monographs. A third of all records also belong to the Crossref category "proceedings-article" which also have a lower share of records with acknowledgements of funding[9].

By looking at some of the records, we also found that some publications are linked to large networking projects. An example is the GrapheneCore1[10] , part of the Graphene Flagship. This

---

[9] For example, a query of records with or without funder information shows that while 23% of journal articles published in 2019 had funding information, this is only the case for 2 % of articles in conference proceedings.

To query the numbers of records with funding information

https://api.crossref.org/types/journal-article/works?rows=0&filter=from-pub-date:2019-01,until-pub-date:2019-12,has-funder:true

https://api.crossref.org/types/proceedings-article/works?rows=0&filter=from-pub-date:2019-01,until-pub-date:2019-12,has-funder:true

for the total remove the filter has-funder:true project reported on SyGMa about 1,000 publications, but most of them acknowledge funding from other FP-funded projects. The same is true for the project "Materials Networking"[11] with over 200 publications. This was a twinning project to enhance the research profile of the Faculty of Chemistry and Pharmacy at Sofia University (FCP-SU) by teaming its researchers to other institutions such as the Materials Science & Metallurgy University of Cambridge, Max-Plank Institute of Polymer Research and Faculty of Chemistry, University of Barcelona. We found that some of the reported publications do not list this funding in acknowledgment but list other funding (which presumably have directly contributed to the results reported) and some seem to be the collaboration of the researchers from the Faculty of Chemistry and Pharmacy at Sofia University and other institutions not listed among the twinning partners.

[10] https://cordis.europa.eu/project/id/696656

This could explain why a large part of the analyzed SyGMa dataset seems to be the only source for those "publication-project" links.

Assuming that they were provided in the reports OpenAIRE gets from SyGMa every 3 months, it is reasonable to deduce that they have not been included in the OpenAIRE Research Graph due to one (or more) of the following reasons:

- A link <publication, project> was passed to OpenAIRE without the DOI. In fact, due to the internal arrangements of management of publication data (project results in general) in EC systems, grant holders can report publications without specifying a DOI. EC internal quality assurance processes may update the record and add the DOI, if it exists. However, the enrichments are not propagated back to SyGMa and, therefore, OpenAIRE is also not notified via the regular updates it gets from SyGMa (as described in Figure 1).
- They did not pass the inclusion criteria by OpenAIRE. Indeed, to avoid the propagation of mistakes that may occur to project managers on the SyGMa platform, OpenAIRE implements several checks including for example comparing the date of publications with the funding period (to detect and filter out publications reported by a project which started years after their publications).
- The DOI of the publication in the SyGMa reports was not well formed and could not be "fixed" with the cleaning rule applied by OpenAIRE, which only removes trailing white spaces and then verifies the DOI is compliant to custom regular expression[12]
- The metadata of a publication could not be retrieved due to technical issues of the Crossref API or – an unlikely event – were not yet available via the Crossref API when the report from SyGMa was processed by OpenAIRE

*3.1. Records in OpenAIRE but not in SyGMa*

We start by noting that, from the dataflow described in Figure 1, one would expect that OpenAIRE has more records than those reported by grant holders. One main reason is that OpenAIRE workflow explicitly records also publications reported after the grant period. Another reason could be the time lag between the publications or deposit in an open access repository (which OpenAIRE indexes) and the reporting period of the project. For the sake of completeness, we also note that a third reason could be the type of publications and the EC H2020 policy on continuous reporting and the Open Access mandate. While grant holders are expected to report on all publication types, the focus in the reporting form is put on "peer-reviewed publications" (for which a link to DOI is explicitly foreseen). Posters, abstracts, reports, and other types of literature research products, which are considered sub-types of publications in OpenAIRE, are usually not peer-reviewed and, therefore, do not fall in the category of publications for which the EC data quality assurance (e.g., linking to external authoritative sources such as Crossref DOI) is performed. This third reason is not relevant for this analysis since we explicitly restricted the comparison to those records with a Crossref DOI.

---

[11] https://cordis.europa.eu/project/id/692146

[12] (10[.][0-9]{4,}[^\s"/<>]*/[^\s"<>]+).

In this comparison, the aim was to check indeed those excess publications (the OpenAIRE Surplus) are genuinely linked to EU funding in a way which can be independently verified.

The workflow adopted is depicted in figure 4.

**Figure 4: Independent verification of records in OpenAIRE but not SyGMa**

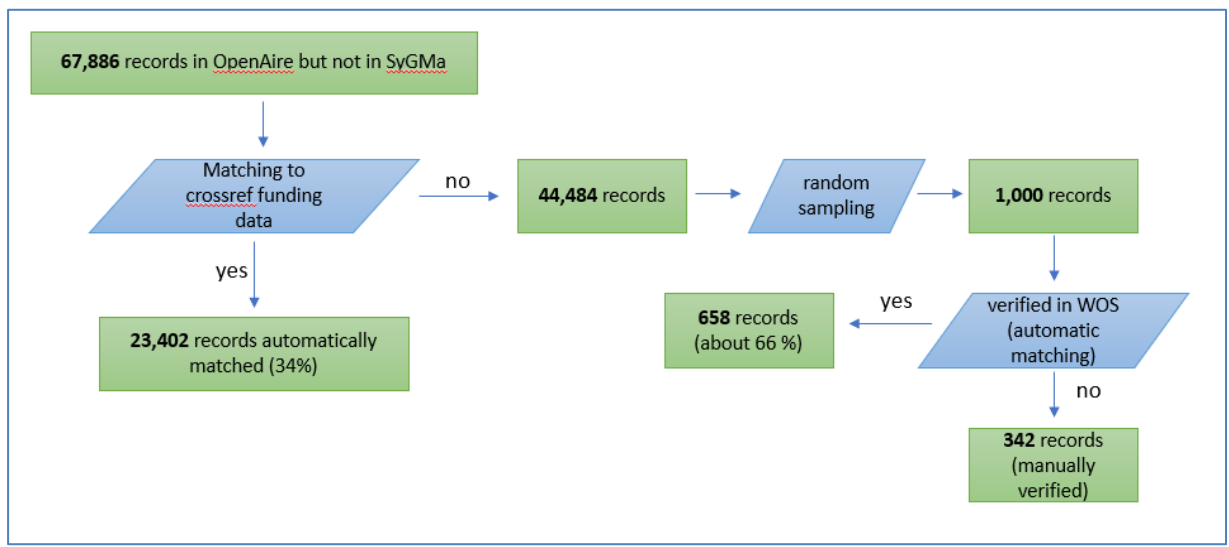

First, we matched the records to Crossref funding data. We used the data dump from January 7, 2021 and could match 23,402 records accounting for 34%. We considered only exact matches between pairs <project, publication doi> in our dataset and <grant number, publication doi> in Crossref funding data. We verified that the funding agency name was the European Union (in different naming).

In a second step, we took among not matched records, a random sample of 1,000 records. We used Clarivate WOS to retrieve both the unstructured funding statement (in the field "fx") and the structured funding data (in the field "fu") and systematically compared the WOS records to the records in our dataset. About 65 % of the records could be matched (658).

The third step consisted in manually verifying the remaining records (342 records). Of those records, 263 links are associated to de-duplicated records. There is a possibility that during this deduplication process, a publication was wrongly attributed to a project. This is the case for example when two – in reality distinct - publications are considered different versions of the same publication and therefore are assigned the same identifier: the funding link propagated to both. OpenAIRE is aware of cases in which this has occurred and is already improving the algorithm for duplicate detection. This does not mean that the link <project, publication> is wrong for all those instances, but that a further check is warranted after the deployment of an improved deduplication algorithm.

For the remaining 79 links, 46 were confirmed as correctly linking the publication to a grant (through acknowledgment or through linking made by institutional repositories, 3 turn out to be data mistake (i.e., the funding link differ from the available acknowledgment / funding statement text), 30 could not be independently verified.

We conclude from this verification, that OpenAIRE "surplus" of publications (i.e., not in SyGMa) could, by and large, be independently verified. The link <project, publication> can be found in acknowledgements/funding statement or in Crossref funding information or in institutional repositories. We also note that about a fifth of records sampled could be found in other systems or in acknowledgements and we could not exclude the related publications have been wrongly attributed to projects. In the next section we make some suggestions on how this can be improved.

## 4. Summary, conclusions, and perspectives

The objective of this analysis was to systematically compare two open datasets of scholarly outputs from EC funded actions: data reported by grant holders (SyGMa data) and data from the OpenAIRE Research Graph. The analysis shows an unexpected set of records (defined as <project,publication> pair) which are recorded in SyGMa but not in OpenAIRE (10% of SyGMa records). We investigated the reason behind this discrepancy and found that it may be due the fact that Crossref links made during the EC internal data quality assurance are not systematically passed on to OpenAIRE. It could also be due to inclusion criteria of OpenAIRE, which enforces strict criteria for example of publications whose release date pre-dates significantly the beginning of the project.

The analysis shows as expected that OpenAIRE has more records than those reported by grant holders. Considering the publications with Crossref DOI, OpenAIRE has 40% more records than the reporting system SyGMa.

The correctness of those records not available in SyGMa has been investigated and the vast majority could be independently verified. This indicates that OpenAIRE offers a more complete picture of the scholarly outputs of EU funded research than the EC reporting system alone[13].

The verification however revealed also some <project,publication> links which could not be independently verified (to some accounts about a fifth of sample of OpenAIRE surplus) which could be the result of data errors made during the deduplication process. OpenAIRE is working on an improved deduplication workflow and a new check will be undertaken after its application. The OpenAIRE team will also continue the checks of the publications involved in the missing links and check if there are common patterns that are not currently supported by the full-text mining algorithm

On overall this exercise also offers some suggestions on how the quality of the data can be improved:

- First of all, it suggested to organize the data exchange between the two systems in such way that changes made in EC systems (related to quality assurance) are made accessible to OpenAIRE. In this context, OpenAIRE should adopt a more "aggressive" cleaning function on DOIs received via the SyGMa reports, in order to minimize the loss of links due to typos that happened at reporting time
- OpenAIRE should put in place a more robust process when querying Crossref, with additional logs and scheduled re-try when the Crossref API does not respond or does not return any metadata. Less strict mapping rules for funding data available in Crossref has shown also to slightly increase the coverage.
- It is also strongly recommended that OpenAIRE provides not only the <project,publication> links in structured form as it does currently but also, where available, the funding acknowledgments/funding statements as extracted from full text. This can help the users detect possible errors and improve the data quality.

---

[13] Although we restricted the analysis to records with Crossref doi, we also note that OpenAIRE has also links to all types of research literature not only peer-reviewed scientific publications, but also slides, presentations, posters, lectures, training materials, etc.

Although we did not investigate the differences between openly available datasets and funding information in commercial databases- an effort we leave for future work -, we observed instances in which Clarivate' Web of Science has funding text and links to EU funding projects that were not present in OpenAIRE. To become the authoritative sources of EU funded scholarly outputs, OpenAIRE should increase the coverage of downloaded full-texts used by the full-text mining algorithm.

We note that the creation of a Research Data Graph of the OpenAIRE magnitude is a complex operation in which errors are inevitable. We believe that analysis like this one can help OpenAIRE and other operators of open data platforms improve the quality of their data. We encourage the community to undertake periodically such quality checks.

**Competing Interests Disclosure**

A.M. Mugabushaka is a policy analyst at the European Research Council Executive Agency (ERCEA). This work was performed while on secondment to the Directorate General Research and Innovation of the European Commission. M. Baglioni & A. Bardi are partly funded by OpenAIRE-Nexus. P. Manghi is CTO of the OpenAIRE A.M.K.E and Technical Manager of OpenAIRE-Nexus. The OpenAIRE-Nexus project has received funding from the European Union's Horizon 2020 research and innovation programme under grant agreement No 101017452. OpenAIRE A.M.K.E is a Non-Profit Partnership (NPP) incorporated under the provisions of Greek Law.